\documentclass[]{pasj02}

\Received{}
\Accepted{}
 
\begin{document} 

\title{ 
Anisotropic quenching beyond $z=1$ and its implications for preprocessing around high-redshift galaxy clusters}

\author{
Makoto \textsc{Ando}\altaffilmark{1,2} \thanks{makoto.ando.astro@gmail.com}, \orcid{0000-0002-4225-4477}
Kazuhiro \textsc{Shimasaku},\altaffilmark{3,4} \orcid{0000-0002-2597-2231}
Kei \textsc{Ito},\altaffilmark{5,6} \orcid{0000-0002-2597-2231}
Takumi S. \textsc{Tanaka},\altaffilmark{3,7,8} \orcid{0009-0003-4742-7060}
Suin \textsc{Matsui} \altaffilmark{3} \orcid{0009-0009-7403-8603}
}%
\altaffiltext{1}{Institute for Cosmic Ray Research, The University of Tokyo, 5-1-5 Kashiwanoha, Kashiwa, Chiba 277-8582, Japan}
\altaffiltext{2}{National Astronomical Observatory of Japan, 2-21-1 Osawa, Mitaka, Tokyo, 181-8588, Japan}

\altaffiltext{3}{Department of Astronomy, Graduate School of Science, The University of Tokyo, 7-3-1 Hongo, Bunkyo-ku, Tokyo 113-0033, Japan}

\altaffiltext{4}{Research Center for the Early Universe, Graduate School of Science, The University of Tokyo, 7-3-1 Hongo, Bunkyo-ku, Tokyo
113-0033, Japan}

\altaffiltext{5}{Cosmic Dawn Center (DAWN), Denmark}

\altaffiltext{6}{DTU Space, Technical University of Denmark, Elektrovej 327 DK2800 Kgs. Lyngby, Denmark}

\altaffiltext{7}{Kavli Institute for the Physics and Mathematics of the Universe (WPI), The University of Tokyo Institutes for Advanced Study, The University of Tokyo, Kashiwa, Chiba 277-8583, Japan}
\altaffiltext{8}{Center for Data-Driven Discovery, Kavli IPMU (WPI), UTIAS, The University of Tokyo, Kashiwa, Chiba 277-8583, Japan}

\email{makoto.ando.astro@gmail.com}


\maketitle

\begin{abstract}
Recent studies have shown that, within galaxy clusters, quenched satellite galaxies tend to be distributed preferentially along the major axis of the central galaxy, dubbed anisotropic quenching. There are various discussions about the origin of this anisotropy: some link it to active galactic nucleus activity in the central galaxy, while others attribute it to the preprocessing of galaxies within large-scale structures outside clusters. However, the definitive cause and its redshift dependence remain unclear. In this study, we investigate anisotropic quenching with 12 spectroscopically confirmed galaxy clusters at $0.9<z<1.4$. We calculate the quiescent satellite galaxy fraction as a function of orientation angle measured from the central galaxy's major axis. Although the statistical significance is modest ($\sim 2\sigma$), we detect anisotropic quenching in the highest redshift ever. To understand the origin of the observed anisotropy, we examine the accretion history of satellite galaxies in a cosmological simulation. We find that, in the $z=1.25$ clusters, the majority of satellite galaxies are recently ($\lesssim 2\,\mathrm{Gyr}$) infalled galaxies. In addition, the orientation angles of satellites are randomized immediately after accretion in $\sim 2\,\mathrm{Gyr}$, suggesting that only recently accreted galaxies contribute to the observed anisotropy. We adopt a semi-analytic approach that combines the accretion history of satellite galaxies with a quenching model based on a delay-then-rapid quenching framework and parameterizes both intrahalo quenching and preprocessing effects. We find that preprocessing is the dominant contributor to quenching and that the quenched fraction attributable to preprocessing is higher along the major axis than along the minor axis by $\sim20\%$, reproducing the observed anisotropic quenching signal. Our results suggest that anisotropic quenching can serve as a novel probe of preprocessing around high-redshift galaxy clusters.
\end{abstract}

\section{Introduction}

The present-day Universe exhibits a complex large-scale structure known as the cosmic web, consisting of voids, filaments, and nodes \citep{de_Lapparent1986,Bond1996}. It is well known that the properties of galaxies depend on the environments in which they reside (e.g., \cite{Butcher1984, Cooper2007, Kodama2007,Peng2010, Wetzel2012, Cybulski2014, Darvish2016, Kawinwanichakij2017,Moutard2018, Chartab2020, Lemaux2022,Nandi2026}). For example, the well-established morphology-density relation shows that early-type galaxies characterized by elliptical morphology, quenched star formation, and old stellar populations tend to reside in denser environments while late-type galaxies with disk-like morphology, active star formation, and young stellar populations are likely to be in less dense environments (e.g., \cite{Dressler1980, Bower1998, Goto2003, Bamford2009}). This galaxy segregation suggests that galaxy evolution is significantly influenced by its surrounding environment.

Galaxy clusters are the most massive ($M_\mathrm{h}\gtrsim 10^{14}\,M_{\odot}$) virialized systems in the Universe (e.g., \cite{Kravtsov_Borgani_2012, Overzier2016}). They host hundreds to thousands of galaxies, as well as hot X-ray-emitting gas, known as the intracluster medium (ICM). Galaxy clusters are also the densest structures in the Universe, located at the nodes of cosmic filaments (e.g., \cite{Kuchner2020,Malavasi2020,Malavasi2022}), and are dominated by quiescent galaxies (QGs), providing a unique laboratory for studying the role of dense environments in quenching galaxy star formation.

In galaxy clusters, various quenching mechanisms unique to massive host halos are proposed. These include direct stripping of cold gas from galactic disks through ram-pressure (ram-pressure stripping, RPS: e.g., \cite{Gunn1972}), halting gas replenishment due to removal of the gas reservoir (strangulation: e.g., \cite{Larson1980}), and multiple close encounters with other satellite galaxies (harassment: e.g., \cite{Moore1996, Moore1998}). In addition, a fraction of galaxies accreted into clusters may have already been quenched in larger-scale filamentary structures before infall through environmental processes operating there (preprocessing: \cite{Fujita2004,DeLucia2012,Sarron2019,Donnari2021,Werner2022,Kuchner2022}). Understanding the origin of the quenched population in clusters, therefore, requires disentangling the relative contributions of these different quenching mechanisms.

It is known that the distribution of satellite galaxies in clusters is anisotropic: satellites tend to be aligned to the major axis of the central galaxy (i.e., the brightest cluster galaxy: e.g., \cite{Yang2006, Wang2008, Paz2011,Z.Zhang2026}). Interestingly, this alignment is more significant for quenched satellites than star-forming ones \citep{Yang2006,Azzaro2007,Huang2016,Rodriguez2022}. Recently, this representative alignment of QGs, dubbed anisotropic quenching or angular conformity, has attracted attention in the field of galaxy quenching pathways, and several possible interpretations of its physical origin have been proposed.

\citet{Martin-Navarro2021} have detected anisotropic quenching in a low-redshift ($z < 0.08$) galaxy group/cluster and compared their results with a cosmological simulation. They have argued that the anisotropy could originate from directional active galactic nucleus (AGN) feedback from the central galaxy, which weakens RPS along the feedback direction (i.e., minor axis of the central galaxy). Such directional AGN feedback is supported by observations \citep{Zhang2022} and simulations \citep{Truong2021}. However, \citet{Karp2023} have presented a counterargument to this interpretation by reproducing similar anisotropy in a simulation that does not include AGN feedback. \citet{Stott2022} has also detected anisotropic quenching for massive clusters at $0.2<z<0.5$ and proposed another scenario in terms of cluster morphology: a cluster halo is elongated along the major axis of the central galaxy, and some environmental effects are more efficient along that direction at fixed cluster-centric radii. More recent work has proposed an alternative explanation: \citet{Stephenson2025} and \citet{Zakharova2025} suggest that anisotropic quenching may instead arise from the preferential infall of preprocessed galaxies from larger-scale filaments. In this picture, the observed anisotropy reflects the anisotropic accretion of quenched populations rather than in-situ quenching within cluster halos. While the physical origin of anisotropic quenching remains debated, distinguishing between the AGN-feedback and preprocessing scenarios would provide important insights into the role of AGN feedback in dense environments and the impact of filaments on galaxy evolution prior to cluster infall.

Another important question about anisotropic quenching is its redshift dependence.
While many works in the literature have mainly focused on the relatively low-redshift Universe (e.g., $0<z<0.5$), \citet{Ando2023} have explored higher redshift using $>5000$ clusters derived from the Hyper-Suprime Cam Subaru Strategic Program (HSC-SSP: \cite{Miyazaki2018, Aihara2022}) and detected anisotropic quenching at $0.25<z<1$. This finding implies that anisotropic quenching occurs universally throughout a wide range of cosmic history and that the underlying mechanism operates even at relatively early times. If the physical origin of anisotropy at $z\gtrsim1$ can be determined, it would place important constraints on quenching pathways in clusters at the relatively early stage of evolution.

In this study, we extend the exploration of anisotropic quenching beyond $z=1$, using the spectroscopically confirmed high-redshift cluster catalog from the Gemini Observations of Galaxies in Rich Early Environments (GOGREEN) survey and the Gemini CLuster Astrophysics Spectroscopic Survey (GCLASS). We report the tentative detection of anisotropic quenching even at $z>1$. Then we use a cosmological simulation to discuss the physical origins of the observed anisotropy and their implications for the role of preprocessing. 

The structure of this paper is as follows. In section~2, we describe the observational data and the selection of the sample. In section~3, we show the detected anisotropic quenching signal beyond $z=1$. In section~4, we conduct a semi-analytic model analysis based on a cosmological simulation and discuss the implications of the observed anisotropy. Overall discussion is in section~5, and section~6 is devoted to a summary and conclusion. Throughout this paper, we assume a flat $\mathrm{\Lambda}$CDM cosmology with $(\Omega_\mathrm{m},\, \Omega_\mathrm{\Lambda},\, h)=(0.3,\, 0.7,\, 0.7)$ and a \citet{Chabrier2003} initial mass function. We use AB magnitudes \citep{Oke1983} and physical distances.

\section{Data and sample}
\subsection{Cluster sample}
We use a high-redshift cluster catalog from the first public data release of the GOGREEN survey and the GCLASS survey \citep{Muzzin2012, Balogh2017, Balogh2021}, which includes 17 spectroscopically confirmed clusters at $0.8<z<1.5$. Of these, 14 clusters are selected from the Spitzer Adaptation of the Red Cluster Sequence (SpARCS) Survey (\citep{Muzzin2009, Wilson2009, Demarco2010}, which are identified as overdensities of red-sequence galaxies \citep{Gladders2000}. Others are discovered by the South Pole Telescope (SPT) survey through the detection of the Sunyaev–Zeldovich (SZ) signature. \citep{Brodwin2010, Foley2011,Stalder2013}. These clusters are observed in multi-wavelength imaging from the near ultraviolet (NUV) to the near infrared (NIR), as well as in optical spectroscopy. In addition, high-resolution NIR images F160W taken by the Hubble Space Telescope (HST) are available for 12 clusters. These HST images are vital to avoid blending issues when measuring the major axes of central galaxies in dense cluster cores. Therefore, we only use these 12 clusters with HST images in our analysis, which are listed in table~\ref{tab:cluster}.

The cluster mass ($M_{200}$) and radius ($R_{200}$) have been reported in \citet{J.Chan2021}, which are obtained from the scaling relation between these quantities and the line-of-sight velocity dispersion of the spectroscopically confirmed cluster members. We summarize these values in table~\ref{tab:cluster}. For SpARCS1613, whose information is not available in \citet{J.Chan2021}, we refer instead to the values presented in \citep{Biviano2021}.

\begin{table*}
  \caption{Cluster properties used in this study}
  \begin{center}
  \begin{tabular}{lcccccccc}
    \hline
    Name & Redshift & $\sigma_\mathrm{vel}$ & $R_{200}$ & $M_{200}$ & $K_{s,\mathrm{lim}}$\footnotemark[a] & $\log(M_{*,\mathrm{lim}})$\footnotemark[b] & $N_\mathrm{all}$ & $N_\mathrm{q}$ \\
         & & [$\mathrm{km\, s^{-1}}$] & [Mpc] & [$10^{14}\, M_{\odot}$] & [mag] & [$M_{\odot}$] \\[6pt]
    \hline
    SPT0205    & 1.320 & $678 \pm 57$ & $0.79 \pm 0.12$ & $2.22^{+0.89}_{-0.70}$ & $23.25$ & $9.90$  &$55$ & $37$\\ [4pt]
    SPT0546    & 1.067 & $977 \pm 68$ & $1.17 \pm 0.09$ & $6.11^{+1.52}_{-1.30}$ & $23.47$ & $9.64$  &$144$ & $88$\\[4pt]
    SPT2106    & 1.132 & $1055\pm 83$ & $1.23 \pm 0.10$ & $7.65^{+2.02}_{-1.72}$ & $23.19$ & $9.79$  &$160$ & $106$\\[4pt]
    SpARCS0035 & 1.335 & $840 \pm 52$ & $0.93 \pm 0.07$ & $4.14^{+1.00}_{-0.87}$ & $23.81$ & $9.70$  &$99$ & $41$\\[4pt]
    SpARCS0219 & 1.325 & $810 \pm 77$ & $0.79 \pm 0.12$ & $2.51^{+1.33}_{-0.98}$ & $23.27$ & $9.90$  &$44$ & $22$\\[4pt]
    SpARCS0335 & 1.368 & $542 \pm 33$ & $0.67 \pm 0.08$ & $1.60^{+0.65}_{-0.51}$ & $22.91$ & $10.07$ &$37$ & $20$\\[4pt]
    SpARCS1034 & 1.385 & $250 \pm 28$ & $0.24 \pm 0.03$ & $0.08^{+0.03}_{-0.03}$ & $24.22$ & $9.55$  &$18$ & $14$\\[4pt]
    SpARCS1051 & 1.035 & $689 \pm 36$ & $0.88 \pm 0.07$ & $2.51^{+0.65}_{-0.55}$ & $24.17$ & $9.35$  &$59$ & $38$\\[4pt]
    SpARCS1616 & 1.156 & $782 \pm 39$ & $0.92 \pm 0.06$ & $3.29^{+0.69}_{-0.60}$ & $23.76$ & $9.59$  &$95$ & $69$\\[4pt]
    SpARCS1634 & 1.177 & $715 \pm 37$ & $0.85 \pm 0.06$ & $2.66^{+0.60}_{-0.52}$ & $24.01$ & $9.50$  &$70$ & $49$\\[4pt]
    SpARCS1638 & 1.196 & $564 \pm 30$ & $0.71 \pm 0.06$ & $1.52^{+0.42}_{-0.36}$ & $23.94$ & $9.54$  &$42$ & $28$\\[4pt]
    SpARCS1613 & 0.871 & $1185\pm 90$ & $1.59 \pm 0.09$ & $11.1^{+0.25}_{-0.25}$ & $22.55$ & $9.97$  &$211$ & $151$\\
    \hline
  \end{tabular} \label{tab:cluster}
  \end{center}
  \begin{tabnote}
  \footnotemark[a] Limiting magnitude in $K_s$ band (AB mag).  \\ 
  \footnotemark[b] Limiting stellar mass \citep{vanderBurg2013,vanderBurg2020}.   
  \end{tabnote}
\end{table*}

\subsection{Galaxy sample}
We use a photometric redshift (photo-\textit{z}) catalog provided by the GOGREEN/GCLASS data release. 
We mainly use photo-\textit{z}, stellar mass, and rest-frame color information. Photo-\textit{z} is estimated using the \textsc{eazy} code \citep{Brammer2008} based on multi-wavelength (11\text{-}13 bands) photometry from NUV to NIR \citep{Balogh2021, vanderBurg2020}. The other galaxy properties are measured using the \textsc{fast} code \citep{Kriek2015} with the stellar population synthesis model of \citet{Bruzual&Charlot2003}. See \citet{vanderBurg2020} for the details.

\citet{vanderBurg2013,vanderBurg2020} have presented the limiting $K_{s}$-band magnitudes for each cluster at which 80 percent of the injected mock sources to the observed images can be detected. They have also calculated the corresponding limiting stellar mass by assuming relatively old stellar populations (i.e., QGs). The derived mass completeness in $\log(M_{*}/M_{\odot})$ for the 12 clusters ranges from $9.35$ to $10.07$, with a median of $9.67$. With a limited number of cluster samples, we decide not to apply a single stellar mass threshold to our galaxy sample (e.g., $\log(M_{*}/M_{\odot})>10$). Instead, we adopt the individual mass limits for each cluster, enabling us to maximize the sample galaxy size. The limiting magnitudes and stellar masses are summarized in table~\ref{tab:cluster}.

To classify galaxies as QGs and star-forming galaxies (SFGs), we adopt a rest-frame U-V and V-J color criterion \citep{Williams2009, Muzzin2013}: galaxies are QGs if 

\begin{equation}
 \begin{aligned}
    (U-V) > 1.3\ \cap \ (V-J)<1.5) \ \cap \\
     (U-V) > 0.88(V-J)+0.59,
 \end{aligned}
\end{equation}
and others are SFGs. 

\subsection{Cluster membership}

We define cluster membership by considering line-of-sight and projected distances as:
\begin{eqnarray}
    |z_\mathrm{phot}-z_\mathrm{cl}|&<&w_{z} (1+z_\mathrm{cl}), \\
    r_\mathrm{proj} &<& 1.5 R_{200},
\end{eqnarray}
where $z_\mathrm{cl}$ and $w_{z}$ are the cluster redshift and the redshift window size for membership selection, respectively, and $r_\mathrm{proj}$ is the projected distance from the cluster center at $z_\mathrm{cl}$. We use photo-\textit{z} rather than spec-\textit{z} to avoid any biases due to the incompleteness of spectroscopic observation.

In general, the redshift window size should be determined by a compromise between contamination rate and incompleteness in cluster member selection: narrower (wider) windows result in lower (higher) contamination and higher (lower) incompleteness. 
Using the GOGREEN galaxy subsamples with spec-\textit{z}'s, \citet{vanderBurg2020} have compared the number counts of `false negatives' and `false positives' (i.e., galaxies selected as cluster members by spec-\textit{z} but not selected by photo-\textit{z}, and vice versa), finding these two are almost the same at $w_{z}\sim0.08$. In other words, contamination and incompleteness do not significantly affect the net galaxy number counts under this window size. Given these considerations, we adopt $w_{z}\sim 0.05$, a slightly stricter selection criterion than \citet{vanderBurg2020}, to keep our cluster memberships purer. We summarize the number of selected total and quiescent members in table~\ref{tab:cluster}.

\subsection{Measuring position angle}
We first select the galaxies nearest to the cluster center as central galaxies, and others are satellites \citep{Balogh2021}. With the HST F160W images, we measure the position angle (PA) of each central galaxy by fitting a single Sersic profile \citep{Sersic1963} using \textsc{galight}, a Python-based 2D model image fitting package \citep{Ding2021}. \textsc{galight} can not only fit the primary object (i.e., the central galaxy) but also simultaneously fit all detected sources around it (i.e., satellite galaxies) using multiple Sersic models, thereby reducing blending effects in the dense cluster core. The Markov Chain Monte Carlo (MCMC) algorithm determines the best-fit model parameters, and PAs are very accurately determined: the uncertainties are less than $2\degree$.

For each satellite, we then calculate the orientation angle, $\theta$, the projected azimuthal angle relative to the PA of its central galaxy. We note that, in principle, all quantities related to this orientation angle should be symmetric in every $90\degree$. 

\section{Result}
\begin{figure}
	\includegraphics[width=\columnwidth]{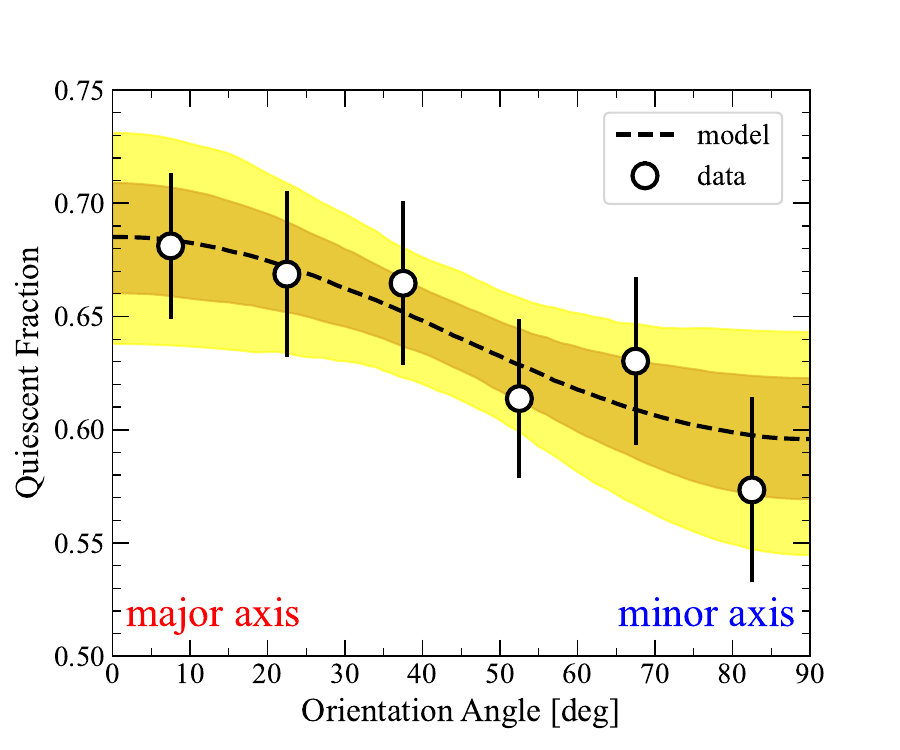}
    \caption{Quiescent fraction ($f_\mathrm{q}$) as a function of the orientation angle measured from the major axis of central galaxies. Black points are $f_\mathrm{q}$ calculated in six angular bins, and error bars are from a binomial distribution. The black dashed line shows the median posterior predictions derived from the MCMC realization, while the dark and light yellow shades show the 68\% and 95\% intervals, respectively.}
    \label{fig:fq_angular}
\end{figure}

We followed the method used in the literature to quantify the anisotropic quenching signal (e.g., \cite{Martin-Navarro2021,Stott2022,Ando2023,Stephenson2025}). Satellite galaxies are divided into arbitrarily chosen angular bins, $\theta_{i}$, and the quiescent fraction is calculated for each bin as:

\begin{equation}
f_{\mathrm{q},i}=\frac{N_{\mathrm{q},i}}{N_{\mathrm{all},i}},
\label{eqn:f_q}
\end{equation}
where $N_{\mathrm{all,}i}$ and $N_{\mathrm{q,}i}$ are the numbers of all galaxies and QGs in the \textit{i}-th angular bin, respectively. The angular dependence is then fitted with a sinusoidal function, and the sinusoid amplitude is used to quantify the degree of anisotropy. We use the cosine function as a fiducial model:

\begin{equation}
f_\mathrm{q ,model}(\theta)=A_\mathrm{q}\cos(2\theta)+f_\mathrm{q,0},
\label{eqn:f_q}
\end{equation}
where $A_\mathrm{q}$ and $f_\mathrm{q,0}$ are the modulation amplitude and the baseline, respectively. Given the small cluster sample size ($N=12$), we adopt six angular bins spanning $15\degree$. The black points in figure~\ref{fig:fq_angular} are the derived quiescent fraction with error bars calculated assuming the binomial distribution. There is a decreasing trend of $f_\mathrm{q}$ from the major axis ($\theta=0\degree$) to the minor axis ($\theta=90\degree$). 

The parameter fitting for equation~\eqref{eqn:f_q} is performed by the MCMC procedure using \textsc{emcee} code \citep{Foreman-Mackey2013}. We adopt uniform prior distributions. The log-likelihood function is:
\begin{equation}
\mathcal{L}=-\frac{1}{2} \sum_{i}\left(\frac{f_{{\mathrm{q}},i}-f_\mathrm{q,model}(\theta_{i})}{\sigma_{\mathrm{q},i}} \right)^{2},
\label{eqn:likelihood_1}
\end{equation}
where $\sigma_{\mathrm{q},i}$ is the error of the quiescent fraction in the \textit{i}-th angular bin. 

\begin{figure}
	\includegraphics[width=\columnwidth]{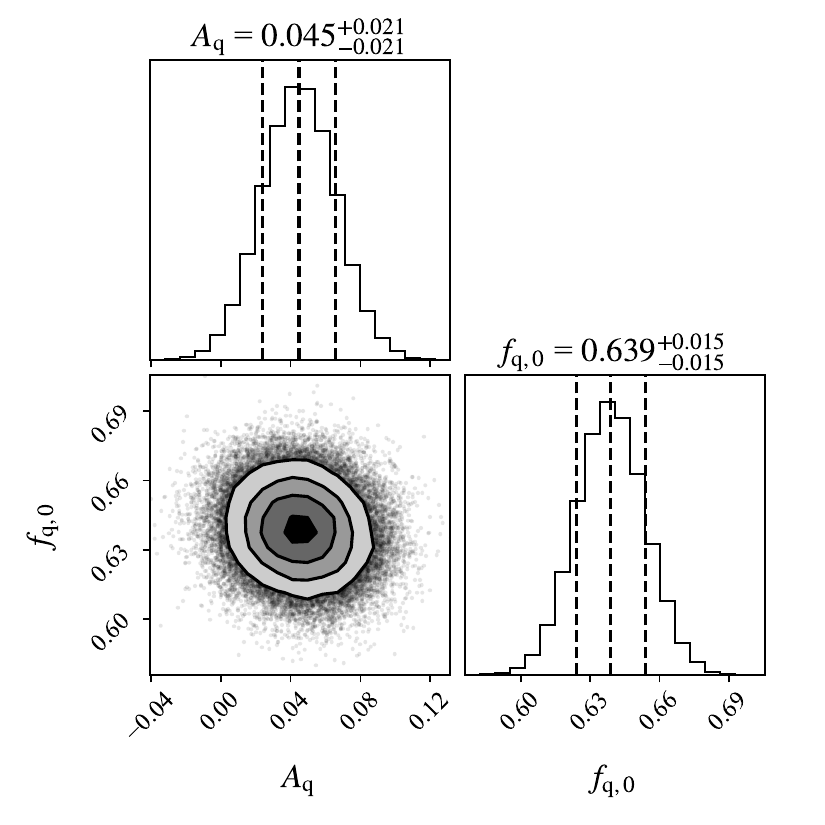}
    \caption{Posterior distributions of $A_\mathrm{q}$ and $f_\mathrm{q,0}$ from the MCMC fitting for the fiducial model. The median value and 68\% credible interval for each parameter are reported at the top.}
    \label{fig:fq_corner}
\end{figure}

In figure~\ref{fig:fq_angular}, we show the median $f_\mathrm{q}$ values (black dashed line) in the MCMC sampling with the 68\% (dark yellow) and 95\% (light yellow) intervals. The cosine function fits the data well, and we derive the median value and 68\% credible interval of each parameter from the posterior distribution (figure~\ref{fig:fq_corner}) as $A_\mathrm{q}=0.045^{+0.021}_{+0.021}$ and $f_\mathrm{q,0}=0.639^{+0.015}_{-0.015}$. We summarize these values in table~\ref{tab:fitting}. The solution of $A_\mathrm{q}=0$ (i.e., no angular dependence) is outside the 98.4\% credible interval, suggesting anisotropic quenching is detected at $2.14\sigma$.

Although this calculation follows a standard approach, it is still possible that using an arbitrary angular bin may affect the anisotropy amplitude. As a cross-check, we also fit a model without binning. by adopting the log-likelihood calculated assuming a binomial distribution as: 

\begin{equation}
\mathcal{L}=\sum_{j}\left[ q_{j}\cdot \log(f_\mathrm{q,model}(\theta_{j})) + (1-q_{j})\cdot 
\log(1-f_\mathrm{q,model}(\theta_{j}))\right],
\label{eqn:likelihood_2}
\end{equation}
where $q_{j}$ takes $1$ if the j-th satellite galaxy with orientation angle $\theta_{j}$ is QG, and $0$ otherwise. The median value and 68\% credible interval for each parameter are $A_\mathrm{q}=0.041^{+0.021}_{-0.021}$ and $f_\mathrm{q,0}=0.639^{+0.015}_{-0.015}$, respectively. The detection significance is $1.95\sigma$. 

Regardless of the calculation method, we derive the non-zero amplitudes of anisotropic quenching. As a complementary check, we fit the data points with a constant value (i.e., an angular-independent) model, obtaining $f_\mathrm{q,0}=0.643^{+0.018}_{-0.019}$. To examine whether the cosine function model is better to fit the data, we calculate the Akaike Information Criterion (AIC: \cite{Akaike1974}) for each model:
\begin{equation}
    \mathrm{AIC_{model}} = -2\ln \mathcal{L}_\mathrm{model} +2k,
\end{equation}
where $\mathcal{L}_\mathrm{model}$ is a likelihood under a given model, and $k$ is the number of parameter of the model. We then derive their difference of $\Delta \mathrm{AIC}=\mathrm{AIC_{fiducial}}-\mathrm{AIC_{const}}=-2.58$, suggesting the fiducial cosine model is mildly favored. Based on these pieces of evidence, we conclude that the signature of anisotropic quenching in the clusters at $z>1$ is confirmed for the first time, although still tentative ($ 2\sigma$ level).

\begin{table}
  \caption{The best-fit parameters of model fit by MCMC}
  \begin{center}
  \begin{tabular}{lcccc}
    \hline
    model & $\mathcal{L}$ & $A_\mathrm{q}$ & $f_\mathrm{q,0}$ & $\sigma$\\[4pt]
    fiducial & eq.~\eqref{eqn:likelihood_1} & $0.045^{+0.021}_{-0.021}$ & $0.639^{+0.015}_{-0.015}$ & $2.14$ \\ [4pt]
    no binning & eq.~\eqref{eqn:likelihood_2} & $0.041^{+0.021}_{-0.021}$ & $0.639^{+0.015}_{-0.015}$ & $1.95$ \\ [4pt]
    constarnt & eq.~\eqref{eqn:likelihood_1} & - & $0.643^{+0.018}_{-0.019}$ & - \\
    \hline
  \end{tabular} \label{tab:fitting}
  \end{center}
\end{table}

\section{Origin of anisotropic quenching}

There are two major possible origins of anisotropic quenching. One is an intrahalo quenching scenario presented by \cite{Martin-Navarro2021}. In this scenario, satellites are less likely to be quenched along the minor axis: AGN feedback from a central galaxy preferentially blows out the ICM along this direction, making quenching driven by RPS less effective. The other is the preprocessing effect \citep{Stephenson2025,Zakharova2025}. Clusters are frequently connected to large-scale filamentary structures where galaxies are more frequently quenched than in the field, and the major axis of the central galaxy and such filaments are likely to be aligned \citep{Rodriguez2022}, causing the anisotropic distribution of QGs in a cluster. Indeed, the anisotropic quenching signature has been reported beyond the cluster virial radius \citep{Stephenson2025}. Since anisotropic quenching beyond $z=1$ is confirmed in this study, investigating its origin may have significant implications for understanding the role of AGN feedback, RPS by the ICM, and preprocessing in the emergence of the quenched populations in high-redshift clusters. 

To understand the physical origin of anisotropic quenching, cosmological simulations are useful because they provide the accretion histories of cluster-member galaxies. Here we use the IllustrisTNG (TNG), a state-of-the-art cosmological simulation \citep{Nelson2019} to examine where and when galaxies are quenched around clusters. We first describe the simulation data in section~\ref{sec:tng}. We then show how satellites' orientation angles are changed after accretion in section~\ref{sec:orienation_angle}. Finally, in section~\ref{sec:toymodel}, we perform semi-analytic modeling of the observed anisotropic quenching signal to infer the origin of this phenomenon at $z\sim1$.

\subsection{Simulation data}
\label{sec:tng}
TNG is a series of cosmological magnetohydrodynamical simulation suites with various baryon physics implemented: metal-line radiative cooling, stochastic star formation, stellar population evolution with associated chemical enrichment and mass loss, stellar feedback, seeding and growth of supermassive black hole and its feedback \citep{Weinberger2017,Pillepich2018a}. There exist simulation runs with different box sizes and resolutions. We focus on the TNG300-1 (TNG300) run, which has the largest box size with a side length of $300\,\mathrm{comoving\ Mpc}$ to include as many massive halos as possible in the analysis.  

In TNG, a friends-of-friends algorithm (FoF: \cite{Springel2001} is adopted to identify groups (halos), and the \textsc{Subfind} \citep{Springel2001} is run to detect substructures in each FoF group as subhalos (galaxies). \textsc{Subfind} defines central galaxies as those containing the most gravitationally bound particle of the parent FoF group. In addition, \textsc{Sublink} \citep{Rodriguez-Gomez2015} is used to link subhalos in different snapshots and generate merger trees (i.e., histories of halo assembly).

From the group catalog, we first select massive halos of $\log(M_\mathrm{200}/M_{\odot})>13.5$ at $z=1.25$ (snapshot \#44) and $\log(M_\mathrm{200}/M_{\odot})>14$ at $z=0$ (snapshot \#99). To ensure a sufficient number of groups for statistical analysis, we adopt the halo mass threshold for the $z=1.25$ snapshot slightly lower than the typical cluster mass regime (e.g., $\log(M_\mathrm{200}/M_{\odot})>14$). Among subhalos belonging to each massive group, we restrict our analysis to those with stellar masses of $\log(M_\mathrm{*}/M_{\odot})>10$ at given redshifts ($z=0$ or $z=1.25$). We have 307 (280) groups and 8104 (10500) galaxies at $z=1.25$ ($z=0$) snapshot. 

To derive the 3-D positions of progenitors of group members at a given snapshot, we trace back merger trees of each galaxy along the main branch. However, merger trees only record the 3-D position at snapshots taken at typical intervals of $150\,\mathrm{Myr}$. Since velocity dispersion in a massive cluster is $>1000\,\mathrm{km\,s^{-1}}$, we may miss sudden changes in the orientation angle of satellites, especially near the cluster centers. Therefore, we perform a cubic Hermite spline (cf. \cite{Patton2024}) to interpolate the object's motion from their 3-D positions and velocities. In the end, the motion of each satellite is traced back over $5\,\mathrm{Gyr}$ with a time step of $10\, \mathrm{Myr}$.

Since observation uses projected position, the orientation angles of satellite subhalos are measured in the x-y, y-z, and z-x planes, tripling the effective sample size. The PA of a central subhalo on the \textit{i-j} plane ($i,j=\{x,y,z\}$) is computed from the inertia tensor $I_{ij}$ using stellar particles associated with that galaxy:
\begin{eqnarray}
    I_{ij}&=&\sum_{n} m_{n} x_{n}^{i}x_{n}^{j}, \\
    \mathrm{PA}_{ij}&=&\frac{1}{2}\arctan\left(\frac{2I_{ij}}{I_{ii}-I_{jj}}\right),
\end{eqnarray}
where $m_{n}$ is the mass of the \textit{n}-th stellar particle, and $x_{n}^{i(j)}$ is its position on the coordinate \textit{i (j)} relative to the subhalo center. Not to include stars extended far beyond the subhalo main body (i.e., intracluster light), stellar particles beyond the minimum of $2r_{*,1/2}$ and $50\,\mathrm{kpc}$ are excluded, where $r_{*,1/2}$ is a half stellar mass radius of a given subhalo. 

\begin{figure}
	\includegraphics[width=\columnwidth]{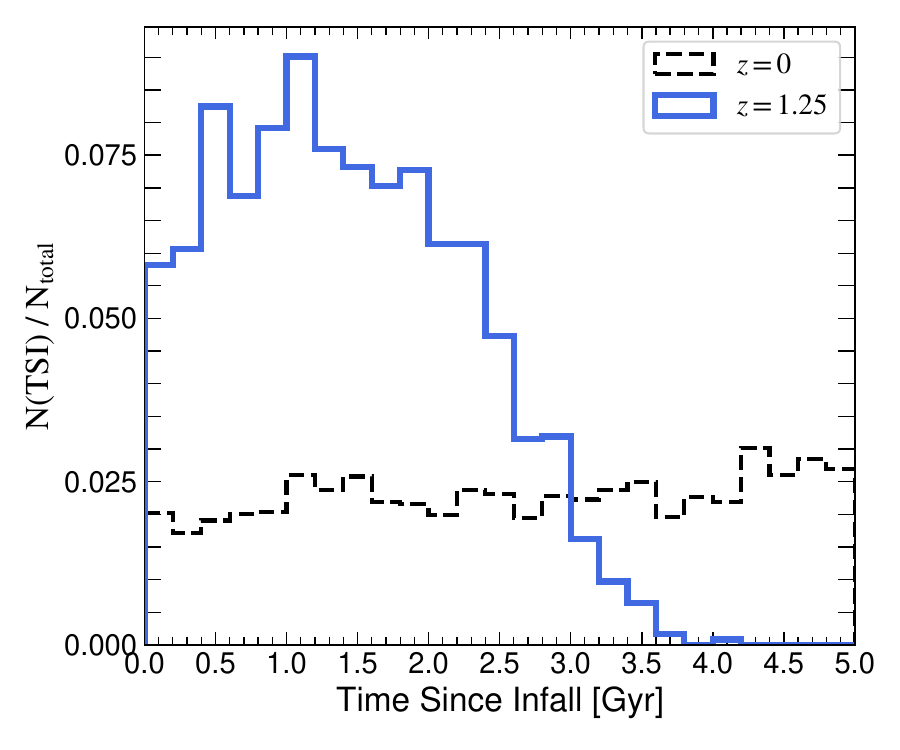}
    \caption{The normalized histograms of the TSI of the satellite galaxies of TNG300 groups at $z=0$ (black dashed line) and $z=1.25$ (blue solid line) snapshots. Most satellites of $z=1.25$ groups are recent infallers with $\mathrm{TSI}\lesssim2\,\mathrm{Gyr}$, and only a small fraction of galaxies have $\mathrm{TSI}>3\,\mathrm{Gyr}$. Note that the sum of the black histogram is not unity since one third of the satellites have $\mathrm{TSI}>5\,\mathrm{Gyr}$ at $z=0$ snapshot, and they are not shown in this plot.}
    \label{fig:TSI_hist}
\end{figure}

Finally, for each satellite, we calculate the time since it first reached $R_{200}$ of its host group (hereafter, time since infall, or TSI). This quantity is commonly used as a measure of how long satellites have been subjected to the extreme environmental effects within a cluster halo (e.g., \cite{Rhee2017}). In Figure~\ref{fig:TSI_hist}, we show the normalized distribution of TSI at $z=0$ and $z=1.25$ snapshots. At $z=0$ snapshot, the satellites have a flat TSI distribution, and one third of them exhibit $\mathrm{TSI}>5\,\mathrm{Gyr}$, which are not shown in the plot. In contrast, those at $z=1.25$ snapshot are skewed toward shorter TSI ($\mathrm{TSI}\lesssim2\,\mathrm{Gyr}$), suggesting that our higher-redshift groups in TNG300 are dominated by the recent infalling satellites.

\subsection{Orientation angle}
\label{sec:orienation_angle}

Given large velocity dispersions in massive clusters, the orientation angles of satellite galaxies may change quickly. Therefore, the observed orientation angles of satellites can differ significantly from those at which satellites are under the strong quenching effects. \citet{Ando2023} have assumed this timescale to be $1-2\,\mathrm{Gyr}$, comparable to a cluster crossing time, and constrain quenching timescale to be $\lesssim 1\,\mathrm{Gyr}$ if intrahalo quenching is the main driver of the anisotropy. However, the actual timescale over which the orientation angle changes significantly from its initial value has not been investigated.

\begin{figure}
	\includegraphics[width=\columnwidth]{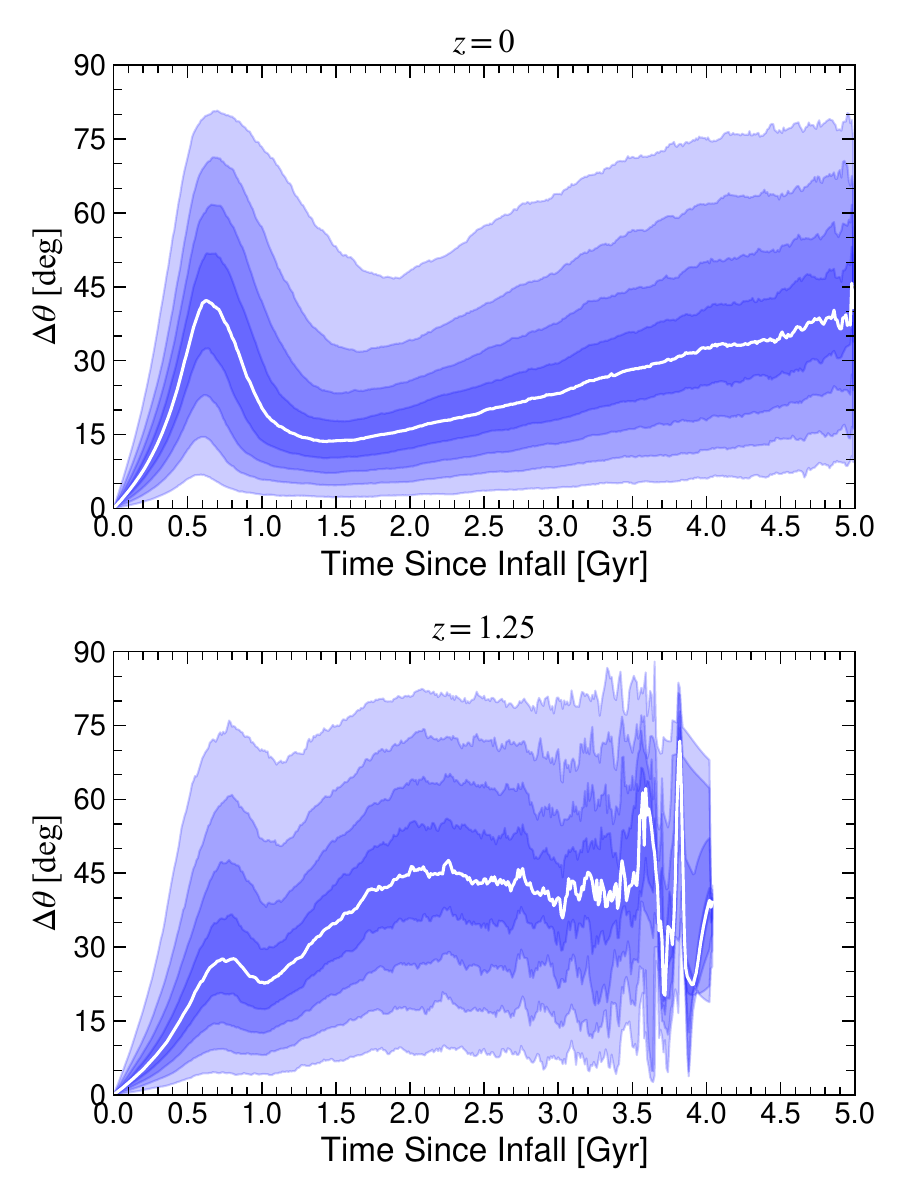}
    \caption{Relative orientation angles at a given TSI against the initial values ($\Delta \theta= \theta(\mathrm{TSI})-\theta(\mathrm{TSI}=0)$) for satellite galaxies in the TNG simulation at $z=0$ (top) and $z=1.25$ (bottom). The white solid line represents the median, and the blue shades indicate the 10th to 90th percentiles, each separated by 10th percentiles.}
    \label{fig:orientation}
\end{figure}

To quantify this timescale, we trace the satellites' orientation angles as a function of TSI. Regardless of snapshots, we always measure orientation angle from the major axes of the central galaxies at $z=1.25$ ($z=0$). In figure~\ref{fig:orientation}, we show the distributions of orientation angles relative to the initial orientation against TSI, $\Delta \theta= \theta(\mathrm{TSI})-\theta(\mathrm{TSI}=0)$, at $z=0$ (top) and $z=1.25$ (bottom). At $z=0$, a clear peak is seen around $0.6\, \mathrm{Gyr}$, probably representing the peri-center passage. After that, the median orientation angle approaches its initial value once ($\mathrm{TSI}\sim1\, \mathrm{Gyr}$) and is gradually randomized in $5\,\mathrm{Gyr}$. This suggests that, if preprocessed galaxies are preferentially accreted onto clusters along the major axis of the central galaxy, their initial orientations are on average preserved at least several $\mathrm{Gyr}$ given the symmetry of the orientation angles in every $90\degree$, making the anisotropic quenching signature. This fact also suggests that the main contributors to the anisotropic quenching signal are recently infalling satellites. In the TNG $z=0$ snapshot, \citet{Zakharova2025} have calculated the amplitude of anisotropic quenching for the young ($\mathrm{TSI}<3\,\mathrm{Gyr}$) and old ($\mathrm{TSI}>3\,\mathrm{Gyr}$) population separately. They have shown that the young population exhibits anisotropy with an amplitude of $0.035\pm0.019$, while the old population shows no anisotropy ($0.002\pm0.009$). This is consistent with our expectations.

At $z=1.25$, the median $\Delta \theta$ shows a peak around $0.7\, \mathrm{Gyr}$, and that peak is much less clear than the $z=0$ case. In addition, the orientation angles are randomized more immediately ($\sim2.0\, \mathrm{Gyr}$). This difference may be due to a difference in cluster crossing time. Since the crossing timescale is approximately proportional to $H(z)^{-1}$, the crossing time at $z\sim1$ is roughly half that at $z=0$. A shorter timescale for randomizing the initial orientation angle compared to $z=0$ may indicate that recently accreted galaxies play a more significant role in arising anisotropic quenching signals. We revisit this point in later sections.

\subsection{Semi-analytic quenching modeling}
\label{sec:toymodel}

To further understand the physical origin of the observed anisotropic quenching, it is necessary to consider not only when and where each galaxy is quenched, but also that the orientation angles of satellites can change over time, as discussed in the previous section. One possible approach is to investigate the quenching mechanisms implemented in hydrodynamical simulations such as TNG300. While this approach has the advantage of fully accounting for complex hydrodynamical processes, it can be challenging to isolate the physical mechanism driving the anisotropic quenching. More importantly, it has been reported that even the latest simulations, including TNG300, fail to reproduce the quiescent fraction observed in the GOGREEN clusters \citep{Kukstas2022}. Therefore, instead of examining the quenching mechanisms implemented in TNG300, we adopt a semi-analytic approach that combines a simple anisotropic quenching model with the satellites’ infall histories obtained from TNG300. Although a semi-analytic approach cannot fully capture the complex hydrodynamical processes, it allows for a more flexible description of the data and facilitates physical interpretation.

We incorporate anisotropic quenching effects into the so-called `delay-then-rapid' quenching model. Our model has four parameters ($f_\mathrm{q,pre}$, $\Delta f_\mathrm{q,pre}$, $\tau_\mathrm{major}$ and $\tau_\mathrm{minor}$) and determines whether a given satellite is quenched depending on its initial orientation angle $\theta_{0}\equiv\theta(\mathrm{TSI = 0})$ and TSI at $z=1.25$. First, we randomly classify galaxies as QGs with probabilities of $f_\mathrm{q,pre}+\Delta f_\mathrm{q,pre}$ and $f_\mathrm{q,pre}$ for those accreted along the major ($0\degree<\theta_{0}<45\degree$) and minor axis ($45\degree<\theta_{0}<90\degree$), respectively; all remaining galaxies are classified as SFGs. This represents quenching that occurs before the cluster infall (i.e., preprocessing), and the anisotropy is implemented as $\Delta f_\mathrm{q,pre}>0$. Remaining star-forming satellites also get `quenched' if they spend time longer than delay times after accretion: $\mathrm{TSI}>\tau_\mathrm{major}$ and $\mathrm{TSI}>\tau_\mathrm{minor}$ for those accreted along the major and minor axis, respectively. This represents quenching within the cluster halo, including any environmental effects such as RPS, starvation, harassment, etc. In this model, $\tau_\mathrm{major}$ and $\tau_\mathrm{minor}$ can take different values if intrahalo quenching works anisotropically. After all satellites are classified into SFG or QG, we calculate the quiescent fraction in arbitrary angular bins on the x-y, y-z, and z-x planes. Note that we do not consider any stellar- or halo-mass dependence for simplicity; this will be implemented in future work.

\begin{figure}
	\includegraphics[width=\columnwidth]{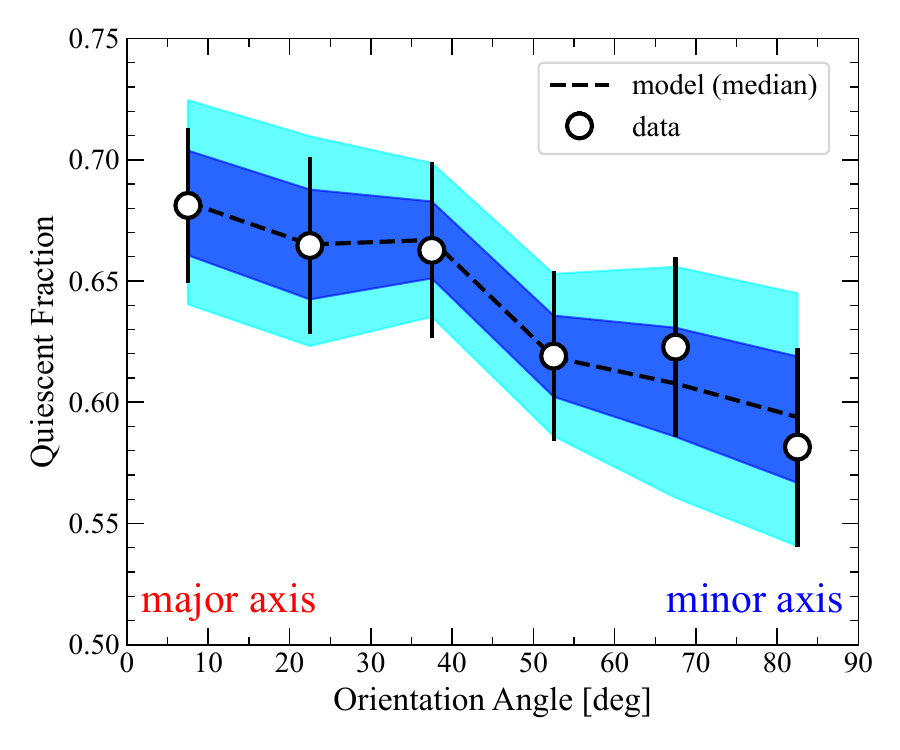}
    \caption{Same as figure\ref{fig:fq_angular} but for the semi-analytic quenching modeling. The black dashed line shows the median posterior predictions derived from the MCMC realization, while the dark and light blue areas represent the 68\% and 95\% intervals, respectively.}
    \label{fig:fq_toymodel}
\end{figure}

\begin{figure}
	\includegraphics[width=\columnwidth]{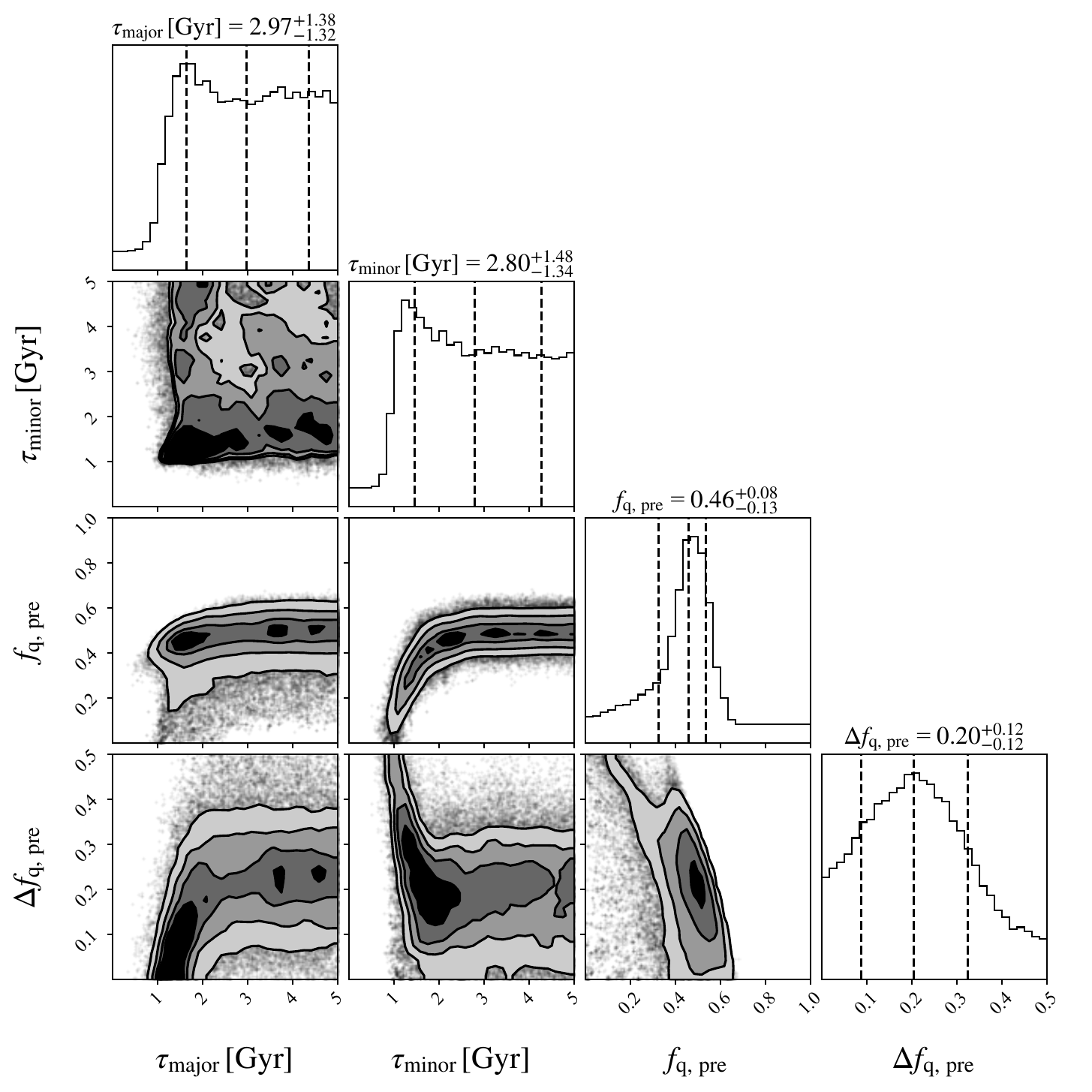}
    \caption{Posterior distributions of four parameters (i.e., $\tau_\mathrm{major}$, $\tau_\mathrm{minor}$, $f_\mathrm{q,pre}$, $\Delta f_\mathrm{q,pre}$) of the semi-analytic quenching model derived from MCMC fitting. The median value and 68\% credible interval for each parameter are reported at the top. Short quenching timescales ($\lesssim 1.5\,\mathrm{Gyr}$) are not preferred while the excess quenching along major axis ($\Delta f_\mathrm{q,pre}$) is significant. }
    \label{fig:toymodel_corner}
\end{figure}

To constrain four parameters, we fit this model to the observationally derived quiescent fraction (see figure~\ref{fig:fq_angular}) via an MCMC procedure adopting the log-likelihood function of equation~\eqref{eqn:likelihood_1} and uniform prior distributions. Figure~\ref{fig:fq_toymodel} shows the derived median posterior predictions of quiescent fraction in the MCMC sampling. Our semi-analytic model (black dashed line) successfully reproduces the observed anisotropy in the quiescent fraction (black points). In table~\ref{tab:toymodel}, we summarize the median and 68\% credible intervals of four model parameters derived from the posterior distributions (Figure~\ref{fig:toymodel_corner}) as fiducial results. In addition, we also report the mode with a 68\% highest probability density interval of the posterior distribution and a representative posterior sample whose parameter set is closest to the median posterior predictions in a least-squares sense. Although the delay times $\tau_\mathrm{major}$ and $\tau_\mathrm{minor}$ are not constrained well, at least the rapid quenching ($\lesssim1.5\,\mathrm{Gyr}$) is not preferred. The baseline quiescent faction $f_\mathrm{q,pre}$ is successfully determined as $0.46^{+0.08}_{-0.13}$. In addition, the excess quiescent fraction along the major axis, $\Delta f_\mathrm{q,pre}$, is also constrained as $0.20^{+0.12}_{-0.12}$. Taking the median value, this suggests that by $\sim 20$ percent more galaxies are likely to be quenched along the major axis than those along the minor axis. This is evidence that preprocessing occurs anisotropically and is a primary driver of anisotropic quenching signals, even at $z>1$.

Conceptually, $f_\mathrm{q,pre}$ and $f_\mathrm{q,pre}+\Delta f_\mathrm{q,pre}$ should be consistent with the quiescent faction observed in the outskirts of the clusters. We estimate the angular-averaged quiescent fraction to be $\langle f_\mathrm{q,pre}\rangle+\langle \Delta f_\mathrm{q,pre}\rangle/2=0.56$, where brackets denote median. Although we do not use the quiescent fraction in the cluster outskirts as a constraint in the fitting, this value is roughly consistent with that in the infalling region (i.e., $1<R/R_{200}<3$) of the GOGREEN clusters \citep{Werner2022}, $f_\mathrm{q}\sim 0.5$ at $\log(M_{*}/M_{\odot})=10\text{-}11$. It is possible that we constrain our quenching model more effectively if we directly use the quiescent fraction in infall regions as a constraint (cf. \cite{Baxter2022}), which is left for future work. One caveat is that the quiescent fraction depends strongly on stellar mass, which we do not account for in our model. To test the validity of our model, including its dependence on stellar mass, a larger cluster sample with deeper imaging data is required. We will address these points in future work.

\begin{figure}
	\includegraphics[width=\columnwidth]{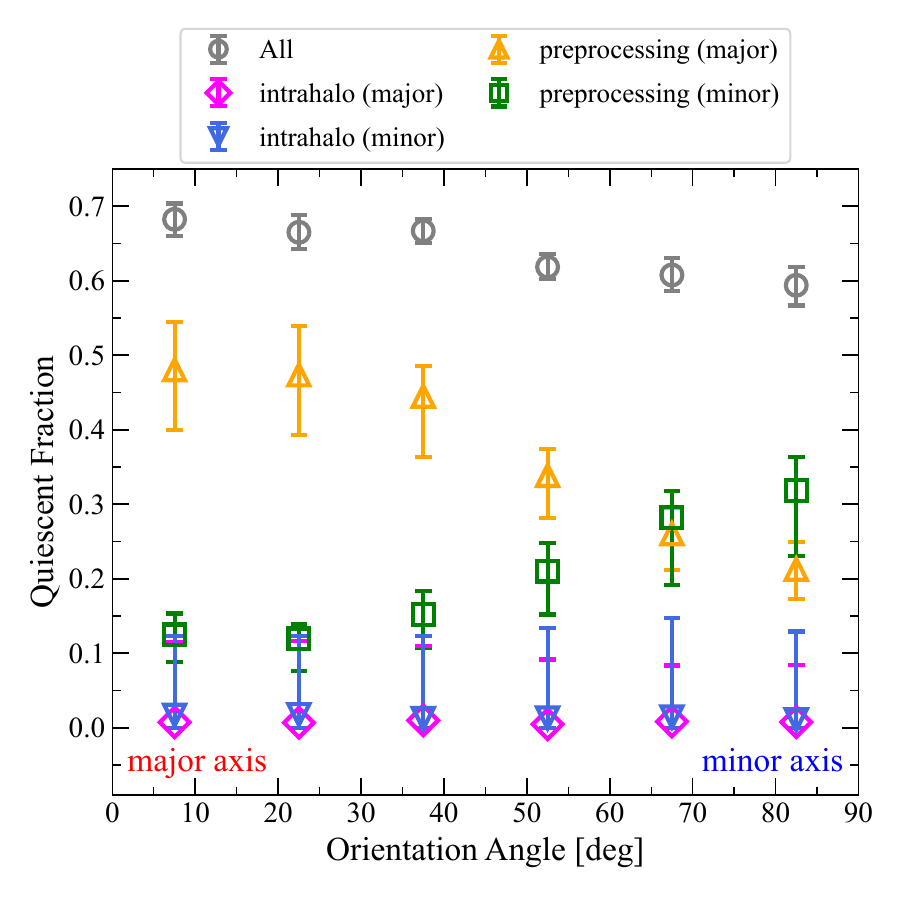}
    \caption{Fraction of galaxies quenched due to four different pathways in our semi-analytic quenching model: preprocessing along major (orange triangle) and minor (green rectangle) axes, and intrahalo quenching along major (magenta diamond) and minor (blue downward triangle) axes.}
    \label{fig:toymodel_cause}
\end{figure}

Finally, we examine which quenching pathways dominate for each `observed' angular bin. We divide the quenched galaxies in the model output into four subcategories according to their quenching path: (1) quenched via preprocessing along the major axis, (2) quenched via preprocessing along the minor axis, (3) intrahalo quenching with a timescale of $\tau_\mathrm{major}$, and (4) intrahalo quenching with a timescale of $\tau_\mathrm{minor}$. Figure~\ref{fig:toymodel_cause} shows the fraction of quenched satellites in each subcategory ($N_{\mathrm{q},i}^{k}/N_{\mathrm{all},i}$, where $k$ represents the subcategories) with their median values and 68\% intervals during the MCMC sampling.

The preprocessed populations that quenched along the major (orange triangles) and minor (green squares) axes are clearly aligned with the axes along which they were accreted. The sum of these two populations accounts for the majority of the total quenched population (gray circles). In contrast, intrahalo quenched populations with $\tau_\mathrm{major}$ (magenta diamonds) and $\tau_\mathrm{minor}$ (blue downward triangles) each contribute at most $\sim10$ percent and show no apparent anisotropy. Given that the rapid quenching is not favored by the MCMC analysis, group members at $z=1.25$, most of which are recent infallers, do not spend sufficient time within the halo to become quenched. Overall, our semi-analytic model indicates that preprocessing is the dominant quenching channel, and that the anisotropy in preprocessing between the major- and minor-axis detections is the primary driver of anisotropic quenching at $z>1$.

\begin{table}
  \caption{Estimated parameters of the semi-analytic model}
  \begin{center}
  \begin{tabular}{ccccc}
    \hline
     Statistic & $\tau_\mathrm{major}$  & $\tau_\mathrm{minor}$ & $f_\mathrm{q,pre}$ & $\Delta f_\mathrm{q,pre}$ \\ 
     & $[\mathrm{Gyr}]$ & $[\mathrm{Gyr}]$ &  &  \\ [4pt]
     median\footnotemark[a] & $2.97^{+1.38}_{-1.32}$ & $2.80^{+1.48}_{-1.34}$ & $0.46^{+0.08}_{-0.13}$ & $0.20^{+0.12}_{-0.12}$ \\[4pt]
     mode\footnotemark[b] & $1.67^{+3.21}_{-0.36}$ & $1.31^{+3.57}_{-0.28}$ & $0.47^{+0.09}_{-0.09}$ & $0.20^{+0.11}_{-0.12}$ \\ [4pt]
     rep.\footnotemark[c] & $1.78$ & $1.60$ & $0.35$ & $0.14$ \\
    \hline
  \end{tabular} \label{tab:toymodel}
  \end{center}
  \begin{tabnote}
  \footnotemark[a] median value and 68\% credible interval.  \\ 
  \footnotemark[b] mode value and 68\% highest probability density interval.  \\
  \footnotemark[c] a representative posterior sample whose model prediction is closest to the median posterior predictions.
  \end{tabnote}
\end{table}

\section{Discussion}

\subsection{Literature comparison}
\label{sec:literature}

\subsubsection{Anisotropic quenching at $z>1$}
\label{sec:literature_1}
Using the GOGREEN/GCLASS cluster sample, we detect anisotropic quenching at $z>1$, extending such measurements to the highest redshift yet achieved. This finding implies that anisotropic quenching has been operating in clusters in their early evolutionary stages. We compare our results with \citet{Ando2023}, who have pioneered the exploration of anisotropic quenching across various redshifts.

\citet{Ando2023} have investigated anisotropic quenching using a photometric redshift catalog of HSC-SSP \citep{Miyazaki2018, Aihara2022, Tanaka2015, Nishizawa2020} and the CAMIRA cluster catalog at $0<z<1.25$ \citep{Oguri2014,Oguri2018}, detecting the anisotropic signature at $>5\sigma$ significance in the $0.25<z<1$ redshift range. They have also examined the anisotropy in the $1<z<1.25$ redshift bin and found the very weak anisotropy with the amplitude of $0.0175\pm0.0082$ ($\sim2.1 \sigma$). \citet{Ando2023} have also argued that they detected the hint of the anisotropy at $z>1$ in the cluster center ($<0.75 R_{200}$) if the quiescent fractions only near the major and minor axes (i.e., $\Delta \theta \pm 15\degree$) are compared. However, due to the weakness of the anisotropy at this redshift bin, they have left the detection of anisotropic quenching at $z>1$ inconclusive. The detection of anisotropic quenching in this work confirms what \citet{Ando2023} have inferred. Compared to them, we detect a stronger anisotropic signal by a factor of $\sim 2.6$, enabling us to overcome significant statistical errors due to a small cluster sample size. However, the origin of the large difference in the observed anisotropy amplitudes is unclear. 

While all the clusters used in this work have been spectroscopically confirmed, only a part of the CAMIRA clusters used in \citet{Ando2023} have been confirmed, which may introduce contamination into the cluster sample from less massive systems and/or filament-like structures aligned with the line-of-sight by chance, possibly weakening the anisotropy. Another possibility is a cluster-to-cluster variation. Compared to \citet{Ando2023}, whose sample cluster size is large ($N=592$), we only use 12 clusters. It is possible that our cluster sample is biased toward those with a large anisotropy by chance. This effect is included in the relatively large uncertainties in the anisotropy amplitude. In addition, the definition of QGs is different: \citet{Ando2023} have defined QGs by the specific star formation rate (sSFR), while we do so by the rest-frame color. Since \citet{Ando2023} have used sSFR inferred from the SED fitting only with optical photometry \citep{Tanaka2018,Nishizawa2020}, their QGs can be contaminated by dusty star-forming galaxies, which may not significantly affect the color-selected QGs in this work \citep{Williams2009}. In any case, a larger confirmed cluster sample at $z > 1$ and the analysis of these clusters in a self-consistent manner with IR data are crucial for conclusively detecting anisotropic quenching in this high-redshift Universe.

\subsubsection{The physical driver of anisotropic quenching}
\label{sec:literature_2}

In the literature, anisotropic quenching has mainly been investigated at $z\lesssim 0.5$ (e.g., \cite{Martin-Navarro2021,Stott2022}). Some recent works have reported that anisotropic quenching occurs not only inside the cluster virial radius but also beyond it. \citet{Stephenson2025} have used 11 massive clusters at $0.2<z<0.5$ from the CLASH survey and detected anisotropic quenching out to $3 R_{200}$ at $>2\sigma$ level. \citet{Zakharova2025} have investigated anisotropic quenching at $z=0$ in TNG. They have detected the anisotropy out to $5 R_{200}$. These studies have explained how anisotropic quenching arises as follows. Clusters are often connected to large-scale cosmic filaments. Because filaments are relatively dense environments and sometimes even host small groups, galaxies within them are quenched more frequently than those outside filaments (e.g., \cite{Sarron2019}). The major axis of a central cluster galaxy tends to be aligned with the filaments, along which these preprocessed galaxies are accreted. Consequently, quiescent satellites are preferentially aligned with the major axis of the central galaxy. As shown in section~\ref{sec:toymodel}, our semi-analytic model infers that preprocessing is a plausible origin of anisotropic quenching at $z>1$, consistent with this explanation. This suggests that preprocessing in filaments along the major axis direction of central galaxies is a main physical driver of anisotropic quenching throughout a wide range of cosmic history.

\citet{Martin-Navarro2021} have originally proposed that the diluted ICM density due to AGN feedback from a central galaxy, and consequently, the weakened RPS causes anisotropic quenching. This scenario is not preferred by our semi-analytic model, which predicts only a limited contribution of intrahalo quenching to anisotropic quenching at $z>1$. In our cluster sample, 3 out of 12 clusters are detected by SZ effects, providing strong evidence for the presence of an ICM. While the other cluster samples are massive enough to host ICM, it is unknown whether they do so. Therefore, there is no clear evidence that supports \citet{Martin-Navarro2021}'s scenario yet. We note that our quenching model incorporates the anisotropy in intrahalo quenching through the initial accretion angles, but a realistic treatment of the ICM distribution would be required to test this scenario more rigorously.

\citet{Ando2023} have supported \citet{Martin-Navarro2021}'s explanation rather than the preprocessing scenario because they have not detected the significant anisotropy at $1<R/R_{200}<2$. This non-detection is opposite to the detection of the anisotropy out to $3 R_{200}$ by \citet{Stephenson2025}. One possible explanation of the non-detection by \citet{Ando2023} is that their sample consists of less massive clusters with $\log(M_\mathrm{h}/M_{\odot})\sim 14.2$ compared to those of \citet{Stephenson2025} with $\log(M_\mathrm{h}/M_{\odot})>14.5$. This may lead to lower contrast between filament galaxies and background/foreground galaxies, resulting in the non-detection of anisotropy beyond $R_{200}$. 

\begin{figure}
	\includegraphics[width=\columnwidth]{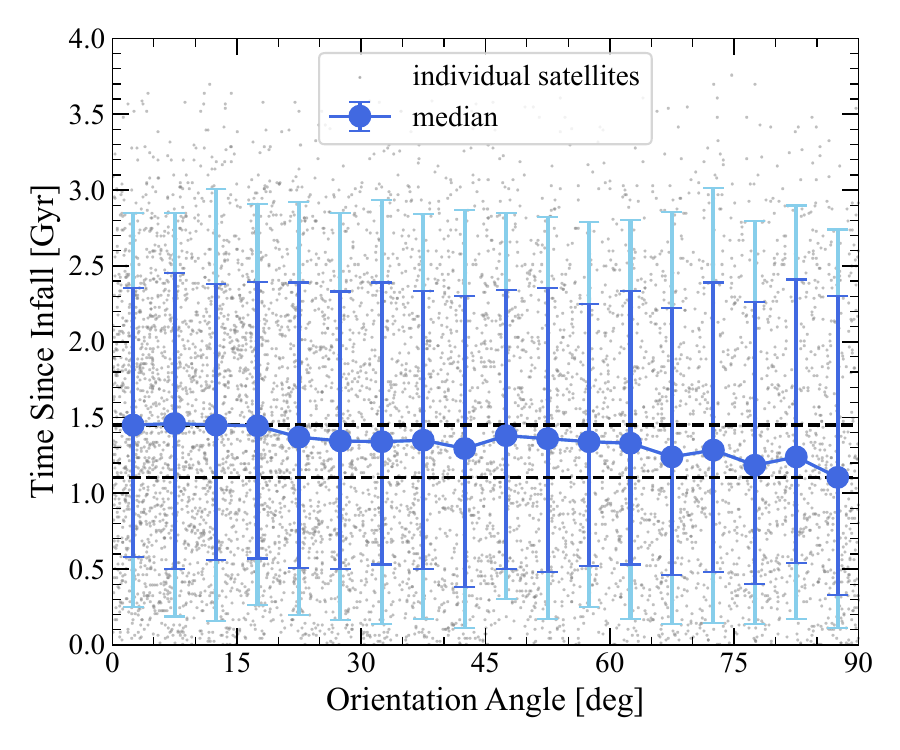}
    \caption{The relation between orientation angle and TSI for the satellites in the TNG300 groups at $z=1.25$. Grey dots represent individual satellites. Blue points indicate their median values and scatters (68\% and 90\% intervals) in bins with $\Delta \theta=5 \,\mathrm{deg}$. Horizontal black dashed lines show the TSI values of the leftmost and rightmost data points, indicating a small difference of $\Delta \mathrm{TSI}=0.345\,\mathrm{Gyr}$.}
    \label{fig:TSI_theta}
\end{figure}

Using the TNG data at the $z=0$ snapshot, \citet{Karp2023} have pointed out that satellites located along the major axis are, on average, accreted onto the halo earlier (i.e., have larger TSI) than those along the minor axis, potentially reproducing the anisotropic quenching signal when a fixed quenching timescale in group halo is assumed. We check whether the same trend persists in our quenching model with the TNG dataset at $z=1.25$. Figure~\ref{fig:TSI_theta} shows the relation between orientation angle and TSI for the satellites. Taking angular bins with $\Delta \theta=5\degree$, we calculate the median values and scatters (68\% and 90\% intervals) of TSI (blue points). The median TSI is almost constant at $\theta<60\degree$ but slightly declines towards the minor axis ($\theta=90\,\mathrm{deg}$), and the difference between the leftmost and rightmost data point is $\Delta \mathrm{TSI}=0.345\,\mathrm{Gyr}$. However, the upper sides of 68\% and 90\% intervals show no clear trend with orientation angles. Considering only the satellites with $\mathrm{TSI}\gtrsim1.5\,\mathrm{Gyr}$ can contribute to the quiescent fraction, this slight difference in TSI distributions is not likely to contribute significantly to anisotropic quenching.

\citet{Werner2022} have pointed out that the majority of massive galaxies in the GOGREEN clusters are likely to be quenched before accretion, either through self-quenching or preprocessing, rather than after accretion. This is consistent with the preprocessing scenario. To the contrary, it is possible that the intrahalo quenching is more effective for less massive galaxies (e.g., $\log(M_{*}/M_{\odot})<10.5$). With deeper imaging data that can detect low-mass galaxies, we can test this hypothesis in the future.

\subsection{Redshift dependence of anisotropic quenching}

By fitting the cosine function to the angular-binned quiescent fraction, we find the amplitude of the anisotropy to be $A_\mathrm{q,0}=0.045\pm0.021$. This value is slightly higher than those reported in the lower-redshift observations (\cite{Martin-Navarro2021,Ando2023}). But to make a fair comparison, we need to take care of halo masses of the samples since it has been suggested that groups hosted by more massive halos show the more significant anisotropy \citep{Martin-Navarro2021,Stott2022,Stephenson2025,Rodriguez2022,Rodriguez2025}. Therefore, when discussing the redshift dependence of anisotropic quenching, group samples with similar halo masses should be compared. Here, we compare our results with those from \citet{Ando2023}, whose sample halo masses roughly match those in this work. 

\citet{Ando2023} have detected anisotropic quenching at $0.25<z<1.0$. Similar to this work, they have used the cosine function to measure the amplitudes of the anisotropic signals. They have derived the amplitudes of the anisotropy in three redshift bins at $5>\sigma$ level: $0.0167\pm0.0048$ ($0.25<z<0.50$), $0.0259\pm0.0030$ ($0.50<z<0.75$), and $0.0303\pm0.0041$ ($0.75<z<1.0$). With our result $0.045\pm0.021$ ($0.9\lesssim z\lesssim1.4$), the amplitudes of the anisotropic signature seem to increase with redshift monotonically. One possible explanation is that the efficiency of the mechanism responsible for anisotropic quenching changes with redshift. For example, the amplitudes of the anisotropy in preprocessing may increase toward higher redshifts. Another possible explanation is that the fraction of recently accreted satellites is higher for higher-redshift clusters. As discussed in section~\ref{sec:orienation_angle}, anisotropic quenching is primarily driven by recently accreted satellites. With a small number of satellites that have spent a long time within the cluster, the fraction of recently accreted galaxies relative to the total population will increase in the high-redshift clusters compared to the lower-redshift ones, resulting in stronger anisotropic signals.

To further discuss the redshift dependence of anisotropic quenching, one approach is to extend the semi-analytic modeling in this work to lower redshifts. If redshift-dependent parameters are implemented in the model, we discuss the redshift evolution of preprocessing in a self-consistent manner.

\section{Summary and conclusion}

In this paper, we investigate anisotropic quenching in the $z\gtrsim1$ clusters observed by the GOGREEN and GCLASS surveys. We measure the quiescent fraction as a function of the orientation angle relative to the major axis of the central galaxy and quantify the anisotropy amplitude using a cosine fit. We derive the anisotropy amplitude to be $A_\mathrm{q}=0.045\pm0.021$, suggesting that anisotropic quenching already occurs even at $z>1$.

To further investigate the physical origin of anisotropic quenching, we use the Illustris TNG cosmological simulation. We then construct a semi-analytic quenching model based on a delay-then-rapid quenching framework, which has four parameters that account for anisotropy in the intrahalo quenching ($\tau_\mathrm{major}$, $\tau_\mathrm{minor}$) and preprocessing ($f_\mathrm{q,pre}$, $\Delta f_\mathrm{q,pre}$). What we find in the simulation analysis and the semi-analytic modeling is as follows:\\

\begin{enumerate}
  \item At the $z=0$ snapshot, the distribution of TSI is almost flat back to $5\,\mathrm{Gyr}$. On the other hand, in $z=1.25$ clusters, most satellite are recently infalled galaxies ($\mathrm{TSI}<2\,\mathrm{Gyr}$), and only a small fraction of satellite have $\mathrm{TSI}>3\,\mathrm{Gyr}$.\\

  \item In the $z=0$ clusters, it takes $\sim5\,\mathrm{Gyr}$ for the orientation angles of satellite galaxies to be randomized. On the other hand, in the $z=1.25$ clusters, orientation angles are more immediately randomized in $\sim2\,\mathrm{Gyr}$ after accretion.\\

  \item Our semi-analytic quenching model well reproduces the observed angular dependence of the quiescent fraction. We find that a short quenching timescale ($\lesssim1.5\,\mathrm{Gyr}$) is not preferred, and that intrahalo quenching is not significant, since most satellites are recent infallers. Instead, preprocessing is a dominant source of quenching.\\
  
  \item The quiescent fraction along the major axis due to preprocessing exceeds that along the minor axis by $\sim20\%$, resulting in the observed anisotropic quenching. This suggests that preprocessing is a main driver of anisotropic quenching in high-redshift clusters.\\

\end{enumerate}

Given the detection of anisotropic quenching beyond $z=1$, this phenomenon is found to be universal across a wide range of cosmic ages. Moreover, if the anisotropic quenching signal reported in this study arises from preprocessing within filaments, this link in turn opens the possibility of using cluster galaxies as a probe of environmental processes in the cosmic web in the early Universe. The large samples of high-redshift galaxy clusters expected from upcoming surveys with the Subaru Prime Focus Spectrograph (PFS), the Rubin Legacy Survey of Space and Time (LSST), Euclid, and the Nancy Grace Roman Space Telescope will enable detailed investigations of environmental effects working in filaments through anisotropic quenching measurements.

\begin{ack}
We appreciate Satoshi Kuriki, Shotaro Akaho, Tsutomu Takeuchi, and Masayuki Tanaka for their helpful discussions. MA acknowledges support from the Japan Society for the Promotion of Science (JSPS) Grant-in-Aid for Scientific Research (26K17202). MA also acknowledges that this work was supported by the Data-Scientist-Type Researcher Training Project of The Graduate University for Advanced Studies, SOKENDAI. TT is supported by JSPS KAKENHI Grant Number JP25KJ0750 and the Forefront Physics and Mathematics Program to Drive Transformation (FoPM), a World-leading Innovative Graduate Study (WINGS) Program at the University of Tokyo.

\end{ack}


\bibliographystyle{plainnat_pasj}
\bibliography{bibtex_26} 

\end{document}